\documentclass{PoS}

\title{Measuremements of the top-quark mass and polarization at the Tevatron}

\ShortTitle{Top-quark mass and polarization at Tevatron}

\author{\speaker{Boris Tuchming}\thanks{On behalf of the CDF and D\O\ collaboration.}\\
    CEA Saclay - Irfu/SPP  \\
    E-mail: \email{tuchming@cea.fr}}

\newcommand{\invisible}[1]{}

\newcommand{\rien}{}

\newcommand{\INVISIBLE}[1]{}

\def\eg{{\it e.g.}}

\newcommand{\formerlongtext}[1]{
}

\newcommand{\dzero}  {D0}

\newcommand{\sqrts}{\ensuremath{\rien{\sqrt{s}}}}

\newcommand{\gev}  {\ensuremath{\mathrm{ GeV}}}
\newcommand{\tev}  {\ensuremath{\mathrm{ TeV}}}

\newcommand{\gevcs}{\ensuremath{\mathrm{ GeV/c^2}}}

\renewcommand{\gevcs}{\gev}

\newcommand{\fbinv}{\ensuremath{\mathrm{ fb^{-1}}}}

\newcommand{\pt}{{\ensuremath{ p_T}}}

\newcommand{\MET}  {\ensuremath{\not{\! \! \! E_{ T}}}}

\newcommand{\etmis}{\MET}

\newcommand{\ppbar}  {\ensuremath{p\bar p}}

\newcommand{\ttbar}  {\ensuremath{t\bar t}}

\newcommand{\BEA}{\begin{eqnarray}}
\newcommand{\EEA}{\end{eqnarray}}

\def\beq  {\begin{equation}}
\def\eeq  {\end{equation}}

\newenvironment{packed_itemize}{
\begin{itemize}
  \setlength{\itemsep}{1pt}
  \setlength{\parskip}{0pt}
  \setlength{\parsep}{0pt}
}{\end{itemize}}

\abstract{We present the most recent and sensitive measurements of the top quark mass, performed by the Tevatron experiments, \dzero\ and CDF, using $\ppbar$ collision data at $\sqrts=1.96~\tev$.
We also present the first measurement of the top quark polarization at Tevatron, obtained by  the \dzero\ Collaboration in the dilepton channels.}

\FullConference{The European Physical Society Conference on High Energy Physics\\
		22--29 July 2015\\
		Vienna, Austria}

\begin{document}

\section{Introduction}

%The top quark  is the heaviest known fundamental particle of the Standard Model (SM) 
%with a mass of~$\simeq 175$~\gevcs, close to the  electroweak breaking scale. 
%This large mass delayed for several years its discovery which occured 20 years
%ago at the $p \bar p$ collider Tevatron~\cite{topdiscovery-D0,topdiscovery-CDF}, during the so called Run I~(1992-1996).
During the Run II~(2001-2011) of Tevatron at $\sqrts=1.96~\tev$, a large amount of $\ppbar$ collisions has been recorded in the \dzero\ and CDF experiments, allowing  to precisely study the top quark properties from  $\ppbar\to\ttbar$ processes. The results presented hereafter rely on the full dataset, approximately ~10~\fbinv\ per experiment, which represents several thousands of observed $\ppbar\to\ttbar$ events.

The top quark  is the heaviest known fundamental particle of the Standard Model (SM) with a mass of~$\simeq 175$~\gevcs, close to the  electroweak breaking scale.
%The top mass is an important ingredient of the SM. 
Precise measurements of its mass allow for consistency checks of the SM and could also determine whether the electroweak vacuum is unstable,  metastable, or stable~\cite{Degrassi:2012ry}. In the following, we report the most recent and accurate top quark mass measurements performed at the Tevatron.

The top quark  has a very short lifetime and decays before it hadronizes. The bare quark properties are transferred to its decay products which allows for their precise measurements.
In the following,
we present the measurement of one of these properties, the top quark polarization.
Since  $\simeq 85\%$ of the \ttbar\ production occurs via $q\bar q$ annihilation in $\ppbar$ collision, the Tevatron has to be viewed as a  $q\bar q$ collider, as opposed to the LHC which is rather a $gg$ collider. Measurements performed at the Tevatron is therefore complementary to those performed at the LHC.

\section{Decay channels}

Within the Standard Model, the top quark decays into a  $W$ boson and a $b$ quark
almost 100\% of the time. The different channels arise from the possible decays of the pair of $W$ bosons:
\begin{packed_itemize}
\item  The ``lepton+jets'' channels ($\simeq 30\%$)
correspond to events where one $W$ decays hadronically and the other  into an electron or a muon.  These channels have a moderate yield and a moderate background arising from  $W$+jets production, $Z$+jets production, or QCD processes. 

\item The ``dilepton'' channels ($\simeq 4.5\%$)
correspond to events where both $W$ decay  into electrons or muons. 
These channels are very pure but have a small yield. The background is due  to  $Z$+jets,
 but also receives contributions from  diboson, $W$+jets, and multijet production.

\item
The ``all jets'' channel ($\simeq 44\%$)
is made of events where both $W$ bosons decay hadronically.
The yield is high, but the background arising from QCD multijet production is very large.

\item  
``tau channels'' ($\simeq 22\%$) arise from events where at least one of the $W$ decays into a $\tau$. As the $\tau$ decays are hard to identify, especially in a hadronic environment as at Tevatron, they are not exploited in the following.

\end{packed_itemize}

Because of the high mass of the top quark,
the decay products have high momenta and large angular separations.
Reconstructing and identifying the production of top quarks demands
reconstruction and identification  of high transverse momenta ($\pt$) electrons, muons, jets, and the measurement of the missing transverse energy (\etmis). 
Good momentum  resolution  for these objects is  required and
the jet energy scale (JES) has also to be known with a good precision.
Eventually, identifying the $b$-jets is  an effective way of improving the purity of the selections.
\section{Top mass measurement}
 To measure the top mass both CDF and \dzero\ have performed 
various analyses, using the three different channels, 
lepton+jets, dilepton, and fully hadronic, and employing different methodology.
The various methods presented below can be classified in three  families:
\begin{packed_itemize}
\item
The template methods consist of choosing  kinematic observables,
creating signal templates for their distributions at different $m_ {top}$ using the MC,
and using a likelihood fit to determine the best signal template.
\item 
The matrix method consists in building an event probability based upon the
proton parton distribution functions, the matrix elements of  $t\bar t$ process, and the transfer functions which relate detector measurements (\eg:  \pt\ of jets ) to the  top quark decay products (\eg:  \pt\ of partons).
Unmeasured quantities are integrated out and a maximum likelihood fit allows to derive the top quark mass.
This method needs also to be calibrated using MC events.
\item The cross-section method derives the top quark  mass from the measured $\ttbar$ production cross-section.
\end{packed_itemize}

When possible, the uncertainty on the measurement due to the JES can be reduced by performing an in-situ calibration. This calibration exploits the $W\to q\bar q'$ decay, which yields the constraint that the mass of the corresponding dijet system should be close to $  m_W=80.4~\gev$.

\subsection{ CDF lepton+jets mass}
This result is based on a  kinematic fit, where three observables are extracted from each event.
The first two observables are obtained from the best (according to the $\chi^2$ of the fit) parton-to-jet assignments: 
$m_{jj}$, the invariant mass corresponding to the $W\to q\bar q'$ decay, and
the top quark mass estimate $m_t$.
The third variables, $m_t^2$, is the top quark mass estimate arising from the second best combination.

Three-dimensional templates are then produced, corresponding to the distributions of the observables. They are smoothed using kernel function methods.
The maximum likelihood fit of the 3d templates to the data simultaneously determines  the top quark mass and the jet energy scale correction factor.
The measurement of the top quark mass achieves a 0.63\% accuracy~\cite{Aaltonen:2012va}:
$ %\begin{eqnarray}
m_t = 172.85 \pm 0.71  \mbox{ (stat+JES)}   \pm 0.85 \mbox{ (syst)}~\gev.
$ %\end{eqnarray}
The dominant sources of uncertainty arise from the residual JES uncertainty (0.52~\gev) and the signal model (0.57~\gev).

\subsection{ CDF all jets mass}
In this channel, the background is very large, with a signal-over-background ratio of $\simeq 1$ for the purest events of the two-$b$-tag sample.
A kinematic fit is employed to extract two observables per event,
a top quark mass estimate,  and a dijet mass estimate.
The measured top quark mass is obtained from a likelihood fit of the 2d templates to the data, together with the JES correction factor.
The top quark mass is measured with  1.1\% precision~\cite{Aaltonen:2014sea}:
$ %\begin{eqnarray}
m_t  = 175.07 \pm 1.19  \mbox{ (stat)}   \pm 1.57 \mbox{ (syst)}~\gev.
$ %\end{eqnarray}
The dominant sources of uncertainty arise from the JES statistical determination (0.97~\gev), the residual JES uncertainty (0.57~\gev) and the trigger efficiency (0.61~\gev).

\subsection{ CDF dilepton mass}
CDF uses an ``hybrid'' method to reduce the JES uncertainty in this channel. The observable extracted from each event is defined as a weighted average:  $m_t^{\rm{hyb}}= w m_t^{\rm{reco}}+ (1-w) m_t^{\rm{alt}}$, where  $m_t^{\rm{reco}}$ is a top quark mass estimate obtained with a neutrino weighting method, $ m_t^{\rm{alt}}$ is an alternate top quark mass estimate, obtained with angles between leptons and jets without the use of jet energy information, and $w=0.6$ is the optimized weight.
The top quark mass is measured with  1.9\% precision~\cite{Aaltonen:2015hta}:
$ %\begin{eqnarray}
m_t = 171.5 \pm 1.9  \mbox{ (stat)}   \pm  2.5\mbox{ (syst)}~\gev.
$ %\end{eqnarray}
The dominant sources of uncertainty arise from the JES (2.2~\gev) and the signal model ( $\simeq 1~\gev$).

%\subsection{CDF combo ??}

\subsection{\dzero\ lepton+jets mass}
This results is based-on the matrix element method, to fit simultaneously the top quark mass and the jet energy scale correction factor.
The top quark mass is measured with  0.43\% precision~\cite{Abazov:2014dpa},
$ %\begin{eqnarray}
m_t = 174.98 \pm 0.58  \mbox{ (stat+JES)}   \pm  0.49\mbox{ (syst)}~\gev,
$ %\end{eqnarray}
which is comparable to the precision of 0.43\% of the Tevatron+LHC combination  in March 2014~\cite{ATLAS:2014wva}.
The dominant sources of uncertainty arise from the residual JES (0.21~\gev) and the signal model ($\simeq 0.3~\gev$).
In Ref.~\cite{Abazov:2015spa}, \dzero\ provides more details and checks on this measurement. For example the $b$-jet JES is tested using the ratio of \pt\ of the $b$-jets to the light jets, and found to be $b-\mbox{JES} =1.008 \pm 0.0195  \mbox{ (stat+JES)}   \pm  0.037\mbox{ (syst)}~\gev$, compatible with 1.% The $b$-jet JES response was also tested using different generators, and was found to be covered by the existing systematic uncertainties on the flavor dependent jet  energy response.

\subsection{\dzero\ dilepton top quark mass}
The \dzero\ experiment has recently updated the measurement in the dilepton channel using the full Run II dataset. The measurement is based on a neutrino weighting method to extract sensitive information from the collision data.
For each event, first a top quark mass hypothesis is fixed, then two values of pseudorapidity for the neutrinos are assumed (using a Gaussian prior), which allows for solving the kinematics and compute a likelihood measuring the agreement between the calculated and the observed \etmis. This likelihood is integrated over the neutrino directions, to obtain a weight distribution as a function of the top quark mass hypothesis. Two observables per event are extracted from this distribution: its width and mean value. The top quark mass measurement consists in a 2d template method based on these two observables. To reduce the impact of JES uncertainties by a factor of $\simeq 4$, the \dzero\ lepton+jets JES correction factor is propagated to this channel.
The top quark mass is measured with  0.9\% precision~\cite{D0conf6463}:
$ %\begin{eqnarray}
m_t = 173.3\pm 1.4  \mbox{ (stat)}   \pm  0.8\mbox{ (syst)}~\gev.
$ %\end{eqnarray}
The dominant sources of uncertainty arise from the statistical uncertainty in the lepton+jets JES determination (0.5~\gev), the residual JES uncertainty (0.4~\gev), and higher orders effect in the signal model (0.3~\gev).
This measurement, has the smallest systematic uncertainty among all dilepton top quark mass measurements performed at hadron colliders.

\subsection{Tevatron combination}
The combination of Tevatron top quark mass measurements has been performed in July 2014 (see Fig.~\ref{figure1}). It has a 0.37\% accuracy~\cite{Tevatron:2014cka}:
$ %\begin{eqnarray}
m_t = 174.343\pm 0.37 \mbox{ (stat)}   \pm  0.52\mbox{ (syst)}~\gev.
$ %\end{eqnarray}
It has to be stressed that such an accuracy is much better than what had been anticipated at the beginning of Run~II.
This combination will be updated in the near future, to include the new \dzero\ dilepton result and minor updates arising from the CDF dilepton and all-hadronic channels.

\subsection{Top quark mass from cross-section}
This method  exploits the fact that the theoretical inclusive $\ttbar$ cross-section  has a strong dependence as a function of the top quark mass, so that the mass is constrained by the experimental measurement of the cross-section.
This method allows for an extraction of the pole mass of the top quark, whereas the direct measurements presented above measure a so called ``MC mass'', which may suffer from a $O(1)$~GeV theoretical uncertainty. 

The cross-section measurement exploiting the full Run~II dataset recorded by \dzero\ has been presented for the first time at the EPS-HEP conference~\cite{Peters}. The measured  cross-section for a top quark mass hypothesis of 172.5~\gev\ is~\cite{D0conf6453}:
$\sigma= 7.73\pm 0.13  \mbox{ (stat)}   \pm  0.55\mbox{ (syst)}~\mbox{pb}.$
The top quark mass is extracted, accounted for the dependence of the measurement as a function of the top quark mass, using the NNLO calculation top++~\cite{top++}:
$ %\begin{eqnarray}
m_t = 169.5 \pm 3.4~\gev.
$ %\end{eqnarray}
\section{Top quark polarization}
\begin{figure}[t]                    
\includegraphics[width=.48\textwidth]{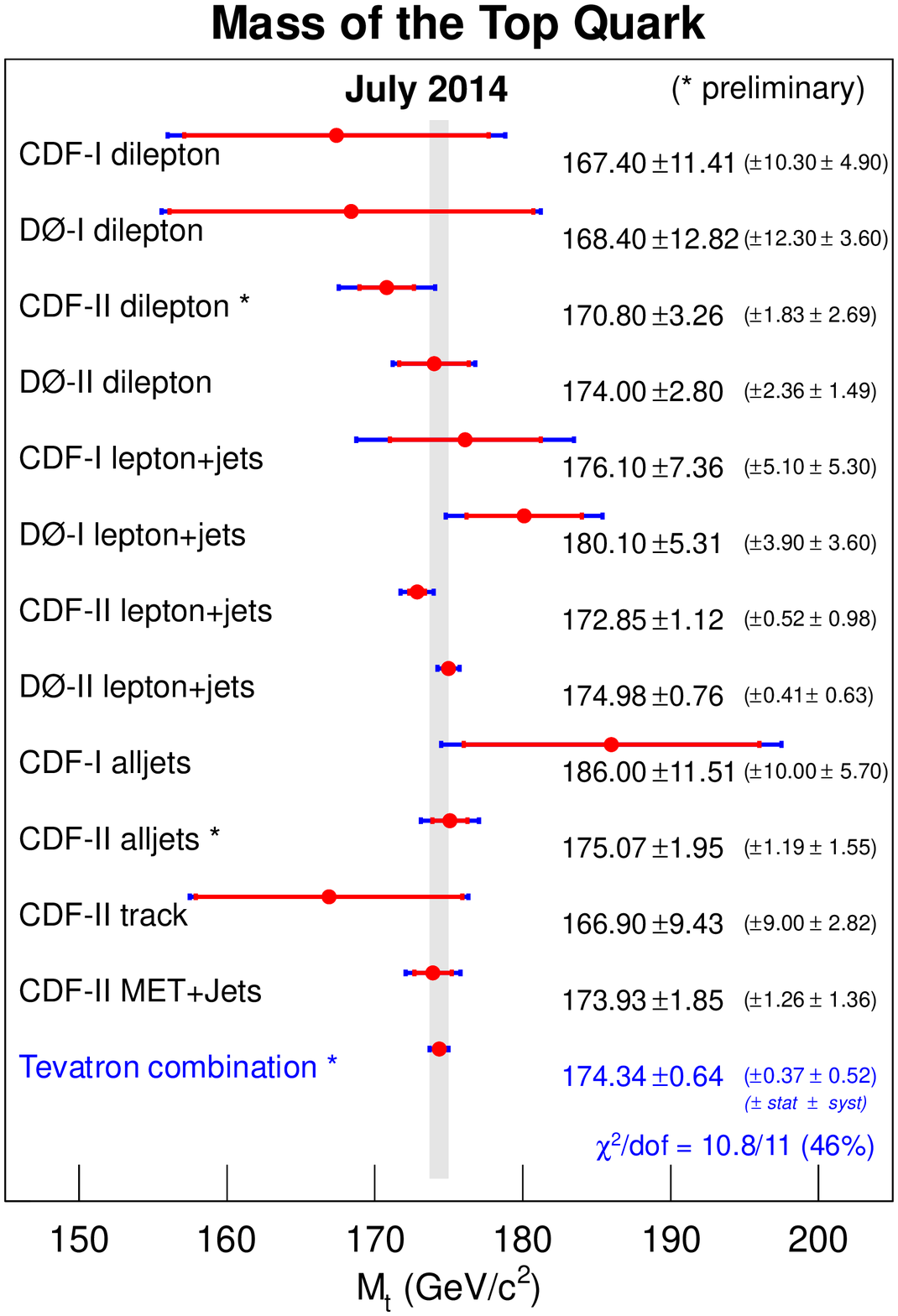}
\includegraphics[width=.52\textwidth]{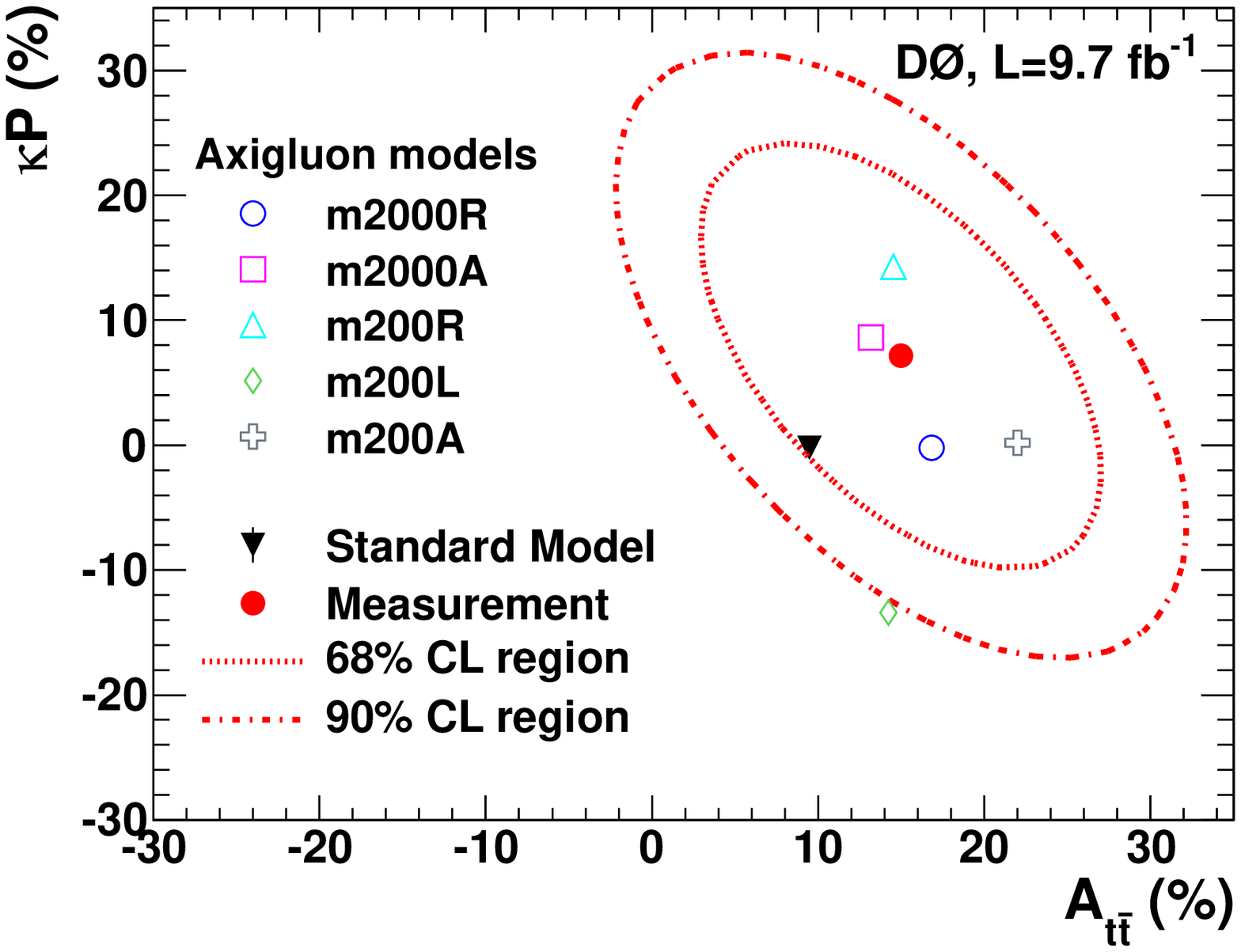}
\caption{Tevatron top quark mass results entering the Tevatron combination (left).  Simultaneous measurement of the forward-backward asymmetry and the top quark polarization in dilepton channel (right).}
\label{figure1}
\end{figure}

%The top quark polarization can be defined and measured in different bases.
%The \dzero\ measurement is performed in 
The \dzero\ polarization measurement is  performed using $\ttbar$ dilepton events in
the so called ``beam basis'', where the main axis is the proton beam boosted in the \ttbar\ rest frame. The simplest observable depending on the polarization is the lepton angle ($\theta^{\pm}$) relative to this axis, where the lepton direction is obtained after successively boosting the lepton in the \ttbar\ rest frame and in the parent top quark rest frame.
The differential angular distribution for $\ell^+$ and $\ell^-$ follows the relation $\def\ddif{d}
{\ddif \sigma}/{\ddif \cos\theta^{\pm}
}=
\frac 1 2 \left( 
1+ \kappa^{\pm} P^{\pm} 
\cos\theta^{\pm}
\right),
$
where $ \kappa$ is the spin analyzing power of the particle ($\simeq 0.99$ within the SM) and $ P$ is the polarization.
Likelihood functions for the $\cos\theta^{\pm}$ observables are obtained
from a matrix element integration method, derived from the method
used for the  lepton+jets top quark mass measurement. The accumulation of likelihood functions over the sample of events is an estimate of the distributions of the cosines.
The polarization is measured from the asymmetry of these distributions, after correcting for acceptance and resolution effects obtained from reweighted MC samples.
This measurement is actually simultaneously performed with the measurement of the forward-backward asymmetry of the \ttbar\ production ($ A^{\ttbar}_{\rm{FB}}$) obtained with the same matrix element method, as correlations occur due to acceptance and resolution effects~\cite{Abazov:2015fna}:
$ %\begin{eqnarray}
 A^{\ttbar}_{\rm{FB}}= (15.0 \pm 6.4 \mbox{ (stat)} \pm 4.9 \mbox{ (syst)})\%,
\kappa P = (7.2  \pm 10.5 \mbox{ (stat)} \pm 4.2 \mbox{ (syst)})\%,   \label{eq:final_result}
$ %\end{eqnarray}
with a correlation of $-56\%$ between the  measurements (see Fig.~\ref{figure1}).
The main systematic uncertainties arise from the hadronization in the signal model ($(3.3\%,1.9\%)$), and the difference in response to different beyond-standard-model benchmarks ($(2.3\%,2.6\%)$).

As a test of the SM, the polarization is extracted constraining $ A^{\ttbar}_{\rm{FB}}$ to its SM prediction
of  $(9.5\pm 0.7)\% $~\cite{Czakon:2014xsa}, which yields:
$ %\begin{eqnarray}
\kappa P = (11.3 \pm 9.1 \mbox{ (stat)}  \pm 1.9 \mbox{ (syst)})\%.
$ %\end{eqnarray}
This first measurement of the polarization at Tevatron agrees with the SM prediction of $(-0.19\pm 0.05)\%$~\cite{Bernreuther:2006vg}.

\section{Summary}
Twenty years after the top quark discovery at the Tevatron, Tevatron data still  provide a valuable insight into top quark physics. Very precise measurements of the top quark mass have been obtained by the CDF and \dzero\ Collaborations. \dzero\ also measures the top quark polarization  for the first time at Tevatron. This measurement is complementary to LHC results, due to the Tevatron \ppbar\ initial states.

\end{document}